\documentclass[nojss, shortnames]{jss}
\usepackage{amsmath}
\usepackage{orcidlink,thumbpdf,lmodern}

\title{Daily-Resolved Lightning Climatology of the Eastern Alpine Region at the Kilometer Scale}
\Plaintitle{Daily Lightning Climatology at the Kilometer Scale}
\Shorttitle{Daily Lightning Climatology at the Kilometer Scale}

\author{Thorsten Simon~\orcidlink{0000-0002-3778-7738}\\Department of Mathematics\\University of Innsbruck
   \And Georg J.~Mayr~\orcidlink{0000-0001-6661-9453}\\Department of Atmospheric\\and Cryospheric Sciences\\University of Innsbruck}

\Plainauthor{T.~Simon and G.~J.~Mayr}

\Abstract{
Lightning flashes are rare albeit hazardous events. Despite this scarcity,
generalized additive models (GAMs) succeed in producing a climatology of
lightning occurrence for the eastern Alps and surrounding lowlands at an
unprecedented resolution of 1\,km$^2$ for each day of April through September
with data from the ALDIS lightning location system. The GAM adds the effects of
seasonality, jaggedness of the terrain, and seasonally varying effects of
elevation and region, thus combining information from analysis cells sharing
similar characteristics. The probability of a cloud-to-ground discharge over
1\,km$^2$ on a given day is typically less than 1\,\% with a rapid increase in
spring, followed by a plateau and a gentler tapering-off in fall. Probabilities
early are lower at high elevations but increase once their snow cover is gone.
Regional patterns of lightning also vary with season with an overall southward
shift later in the year but more complex details. Grid cells with jagged
topography have a higher probability of lightning.} 


\Address{
  Thorsten Simon\\
  Department of Mathematics\\
  University of Innsbruck\\
  Technikerstraße 21a\\
  6020 Innsbruck, Austria\\
  E-mail: \email{Thorsten.Simon@uibk.ac.at}\\
  ORCID: \href{https://orcid.org/0000-0002-3778-7738}{0000-0002-3778-7738}
}

\begin{document}

\section{Introduction}

Lightning affects many fields of research and aspects of everyday life.
Cloud-to-ground lightning strikes may damage equipment and structures such as
wind turbines \citep{montanya2014, becerra2018} and power lines
\citep{cummins1998}, start fires \citep{reineking2010, dowdy2012} and injure or
kill people \citep{ritenour2008, holle2016}. It produces NO$_x$, which  in turn
affects the concentration of greenhouse gases \citep{murray2016}. During warm
season lightning is closely connected with strong convection, which adds flash
floods, large hail and damaging winds as further hazards. Having reliable
climatologies of lightning thus aids the assessment of all these risks and the
understanding of processes associated with strong convection.

Lightning location systems (LLS) measure lightning discharges continuously in
both space and time unlike any other atmospheric measurement system. Detection
efficienies often exceed 90~\% with location accuracies well below 1~km 
\citep[e.g.][]{poelman2016}. However, lightning
is a rare event with typically only a few discharges per square kilometer over
the whole year. Therefore computing climatologies with the simple
``cell-count'' method of counting flashes and dividing it by the overall period
is limited to fairly large cells, where an area times a period constitutes a
``cell''. We will use this spatio-temporal definition of ``cell'' throughout the paper.
Climatologies compiled with the cell-count method typically use cells 
of 10--100~km$^2$ times 1--12 months. 

\citet{diendorfer2008} treated
lightning as a Poisson-process to provide a theoretical limit of the width of
the 90~\% confidence interval of the flash density estimate with the cell-count
method and found that for this width to be $\pm20$~\%, approximately 80 flashes
in a cell over the whole observation period are needed. Two promising
approaches have been used to achieve spatial cell sizes of 1~km$^2$ and
temporal sizes of 1 month or shorter. The first one \citep{bourscheidt2014,
kingfield2017} uses the location uncertainty instead  of the precise locations of 
flashes, which amounts to performing a kernel density estimation around the location of
lightning discharge with an assumed independence between the discharges. 
The second one \citep{simon2017spatio} exploits the
similarity in seasonal, regional and altitudinal characteristics between cells
using the statistical method of generalized additive models to achieve even
smaller cell sizes of 1~km$^2$ times 1 day and additionally extract the
functional dependence of lightning on these underlying characteristics. Their
study was intended as a proof of concept for the suitability of GAMs to achieve
high-resolution lightning climatologies and thus limited to a small region (the
state of Carinthia in Austria) and basic effects of season, elevation and region.

We will use the GAM method to compute lightning climatology for the entire
eastern Alpine region and the surrounding lowlands and include a more
comprehensive set of shared characteristics between cells. To our knowledge it
is the first lightning climatology at a cell resolution of 1~km$^2$ times 1 day
for such a large region. Other climatologies for (parts of) this region
\citep[e.g.][]{feudale2013, wapler2013high, taszarek2019} have a considerably
coarser---either in space and/or in time---cell resolution.

\begin{figure}[t!]
\centering
\includegraphics[width=14cm]{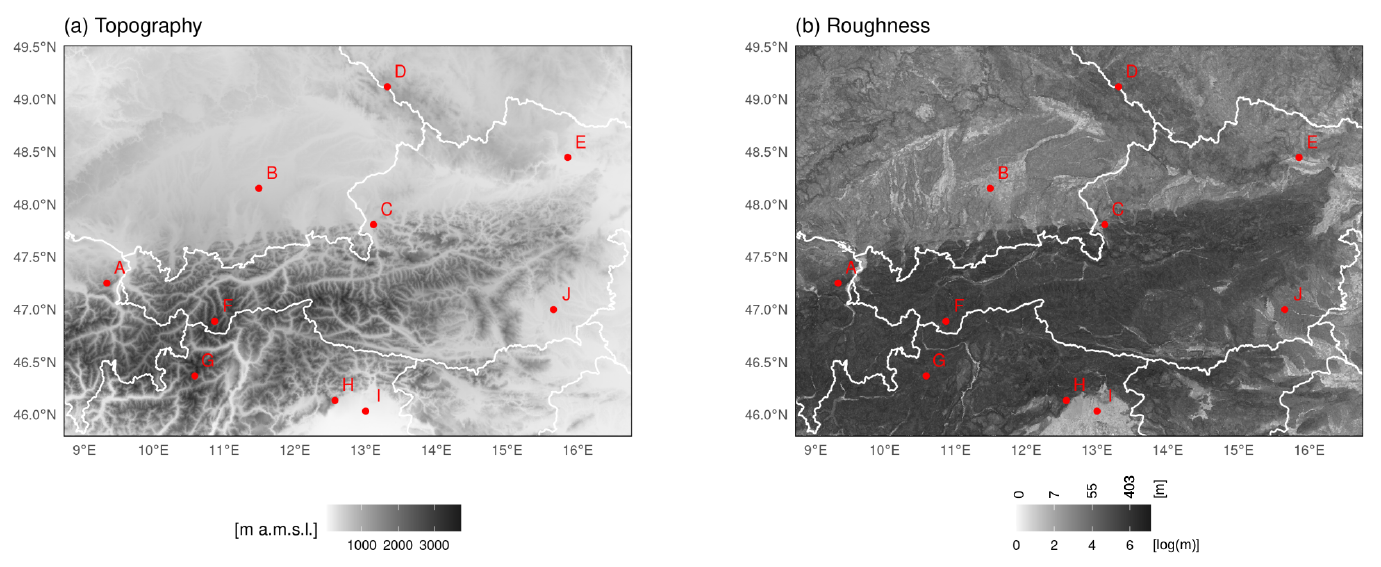}
\caption{Eastern Alps: topography and topographical roughness---the
(logarithm of the) difference between maximum and average elevation in a grid
box. Country borders and locations for which seasonal lightning cycles will be
shown are added. Details on the specified locations A, B, \dots, J are given in
table~\ref{tab:locations}.}
\label{fig:regions}
\end{figure}

\section{Data}\label{sec:data}

The climatology of lightning days is based on lightning detection data from the
Austrian Lightning Detection \& Information System
\citep[ALDIS,\,][]{schulz2016}, complemented by data from the
digital elevation model TanDEM-X \citep{rizzoli2017} to account for topographic
effects.

\subsection{Lightning location data}

For this study we use ALDIS cloud-to-ground flashes for the summer months
April--September. Eleven years of data, 2010--2020, have been on hand.
The ALDIS data are cropped slightly to a domain that covers around a quarter of a
million square kilometers (Fig.~\ref{fig:regions}). The domain includes the
European Eastern Alps and extends northwards to the lower mountain ranges of the
Swabian and Franconian Jura and the Bohemian Forest.

\begin{table}[t]
\centering
\begin{tabular}{rlrrrr}
  \hline
 & Name & Latitude & Longitude & Elevation [m] & Roughness [m] \\ 
  \hline
 A & Saentis & 47.2494 & 9.3433 & 2214 & 271 \\ 
   B & Munich & 48.1513 & 11.4921 & 575 & 13 \\ 
   C & Gaisberg & 47.8056 & 13.1125 & 1063 & 268 \\ 
   D & Bohemian Forest & 49.1182 & 13.3071 & 1235 & 100 \\ 
   E & North of Vienna & 48.4452 & 15.8566 & 268 & 11 \\ 
   F & Central Alps N & 46.8854 & 10.8672 & 3593 & 155 \\ 
   G & Central Alps S & 46.3654 & 10.5872 & 3213 & 270 \\ 
   H & Range NE of Udine & 46.1327 & 12.5675 & 1545 & 127 \\ 
   I & Udine & 46.0301 & 13.0013 & 127 & 5 \\ 
   J & Grazer Becken & 46.9968 & 15.6558 & 479 & 49 \\ 
   \hline
\end{tabular}
\caption{Specifications of locations: The elevation refers to the average
of TanDEM-X data in a hexagonal square kilometer. Roughness is defined by the difference
between maximum and average TanDEM-X elevations in a hexagon. The letters A, B, \dots, J
are used to mark the locations in Fig.~\ref{fig:regions}.}
\label{tab:locations}
\end{table}

The spatio-temporal dimensions of a cell used for our climatology are an
area of 1 square kilometer and a period of 1 day. The shape of the 
1~ square kilometer area is taken to be hexagonal and a day is taken to 
start at 06~UTC, the approximate time of the diurnal lightning minimum. 
The study region has an area of $248\,308$~km$^2$ and the study period 
from April through November contains 182 days, resulting in approximately 
45.2 million spatio-temporal cells for the climatology. For each of these cells 
the number of years is counted in which ``lightning`` occurred, defined as at 
least one cloud-to-ground discharge. Since the data span the eleven years 
2010--2020 this number can in principle lie between 0 amd 11. However, no cell 
had more than five years in which lightning occurred.

\subsection{Digital elevation model}

The digital elevation model TanDEM-X \citep{rizzoli2017} is used to enhance the
data by topographic information. The version used here has a horizontal
resolution of 90~m and is aggregated over the unit square kilometer hexagons by
taking the mean and the maximum. After aggregation $51$ (out of $248\,308$ unit square kilometer cells)
have missing values. They are imputed with the mean of the surrounding values since
all of them are over water bodies such as Lake Constance and Lago di Garda.

The natural logarithm of the mean topography (Fig.~\ref{fig:regions}a) will
enter the statistical climatology model as explainatory variable.  Further, the
natural logarithm of the difference between the maxim and the average topography serves
as proxy for the roughness of the terrain within a
hexagonal square kilometer (Fig.~\ref{fig:regions}b).

\section{Methods}\label{sec:methods}

Each of the 45.2 million spatio-temporal cells---1~square kilometer hexagons spatially 
times 1~day temporally for each day from April through October---is one
data point for our statistical climatology. 
Each cell contains the number of observed
thunderstorm days over the 11~years for this specific day of the year and
hexagonal area. These numbers of observed thunderstorm days $\{0,1,2,\dots,11\}$
can be seen as realizations of Bernoulli's urn problem with replacement. For
each cell, 11~trials are conducted, where each trial has two possible outcomes:
 \textit{lightning} and \textit{no lightning}, respectively.

The outcome of these trials follows a binomial distribution, which is
determined by a probability $\pi$ that lightning occurs. In order to estimate 
the probability $\pi$
of the underlying process, we utilize generalized additive models
\citep[GAMs,][]{wood2017generalized} that incoporate data not only for single
data points but leverage information of data points with similar settings, e.g.\
same region, similar time of the year, similar topographic
conditions.  GAMs have proven their abilities for lightning climatologies in
complex terrains \citep{simon2017spatio}.  The GAM used for this study is set
up as follows,

\begin{equation}\label{eq:gam}
  \text{logit}(\pi) =
  \beta_0 + f_1(\mathtt{yday})
  + f_2(\mathtt{log\,topo, yday})
  + f_3(\mathtt{lon, lat, yday})
  + f_4(\mathtt{log\,roughness}).
\end{equation}

On the left hand side is the $\text{logit}$ transformed probability $\pi$. The
$\text{logit}$ maps the $\pi$ from the probability scale $\left]0,1\right[$
to the real line.  On the right hand side is the additive predictor that
combines the intercept $\beta_0$ of the regression and multiple, potentially non-linear
terms $f_\star$.

\begin{itemize}
\item $f_1(\mathtt{yday})$ is a function of the day of the year and thus
    represents a baseline annual cycle valid for all locations of the domain.
\item $f_2(\mathtt{log\,topo, yday})$ is a function of the log-topography
    and the day of the year. Thus this term allows for deviations of the annual
    cycle from its baseline conditioned on the topography.
\item $f_3(\mathtt{lon, lat, yday})$ additionally allows deviations from the
    baseline annual cycle conditioned on the geographical location.
\item $f_4(\mathtt{log\,roughness})$ adds a correction for the roughness of the
    topography.
\end{itemize}

With the GAM all these functions $f_\star$ are modelled using regression
splines such as cubic regression splines and thin plate regression splines.
The spline bases for functions with two ($f_2$) or three ($f_3$) covariates
are set up using the tensor product of the univaraite spline bases. For the
technical details on splines within GAMs the reader is refered to
\citet{wood2017generalized}.

This setup leads to approximately $2\,500$ regression coefficients of
the GAM. These coefficients are estimated via penalized
maximum likelihood.  Here the amount of smoothing of the potentially non-linear
functions $f_\star$ is determined by generalized cross-validation as implemented
in the \textbf{mgcv} extension for \textsf{R}.
Estimating such a flexible regression model for the
large data set on hand with $45\,192\,046~\text{data\,points}$
is feasible with a fitting algorithm for giga data \citep[implemented in the
\textbf{mgcv} extension for \textsf{R}]{wood2017gigadata}.

\section{Results}\label{sec:results}

\begin{figure}[h!]
\centering
\includegraphics[width=10.7cm]{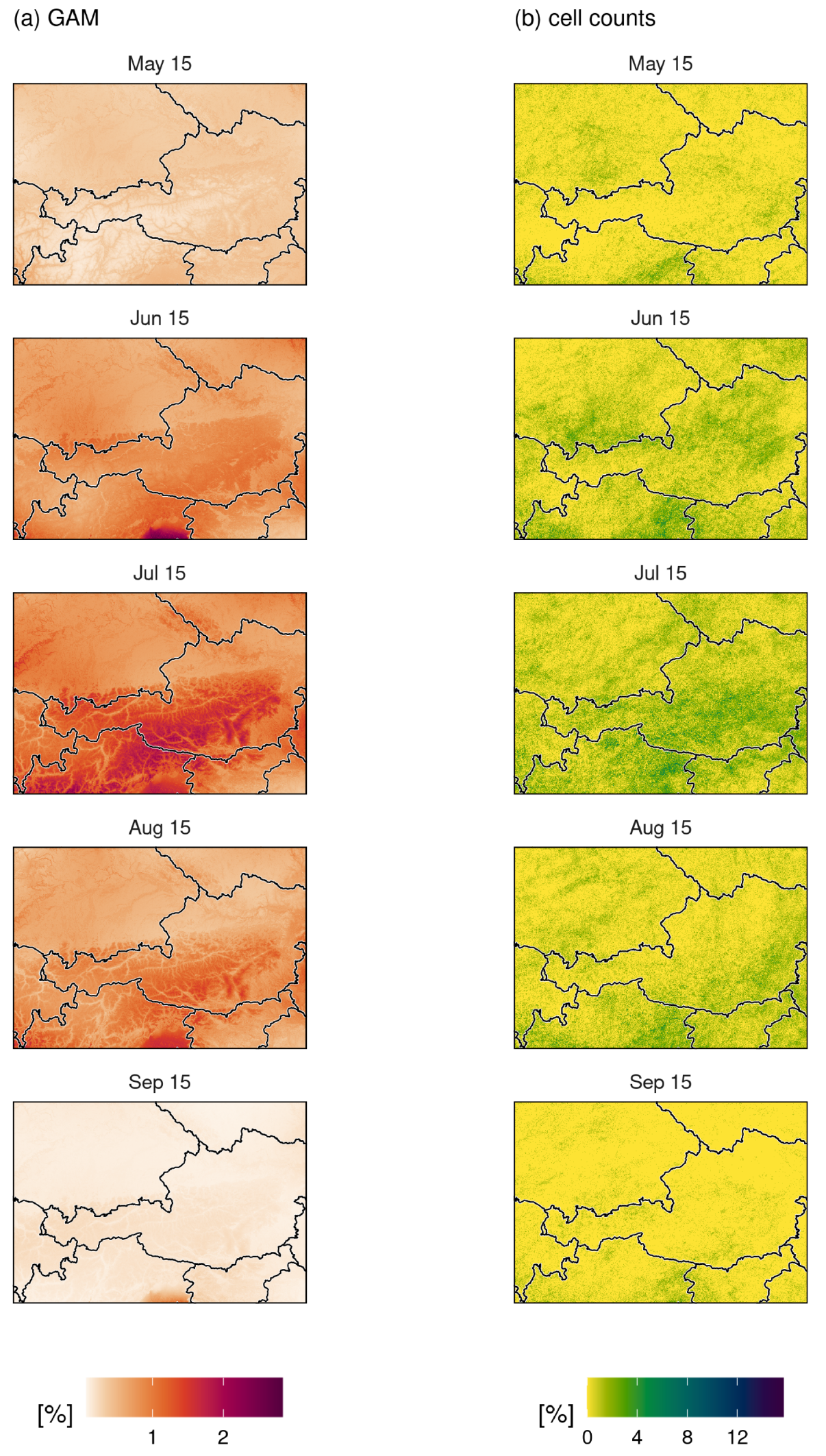}
\caption{Climatology maps for selected days. a. based on the GAM method, and
         b. based on the cell-count methods.}
\label{fig:maps}
\end{figure}

Lightning in the eastern Alps is rare.
Only 5.47~\% of the 45.2 million spatio-temporal cells of 1~km$^2$ times
1 day in the 182 days from months April through September had lightning occur 
in at least one of the eleven years 2010--2020. Fig.
\ref{fig:maps}a shows that the maximum probability for lightning to strike
within a grid box of 1 square kilometer on a particular day barely exceeds
2~\%. The maps for day 15 of each month demonstrate that the probability of
lightning waxes from spring through the end of July and then wanes into autumn.
In spring (May 15, top) the likelihood for lightning strikes north of the Alps
is regionally uniform below about 0.5~\%. The probability along the northern
Alpine rim slightly exceeds these values. In the Alps, valleys and mountain
ranges stand out (cf. Fig. \ref{fig:regions}, top) with probabilities in
valleys considerably higher than over mountain ranges, which are still mostly
snow-covered at this time of the year. By mid-June these differences blur and
valleys are almost no longer distinguishable from mountain ranges. By now,
lightning in the higher terrain north of the Alps strikes more readily than in
the surrounding flat lands. Differences south of the Alps are even starker. The
flat land around location I in Fig.~\ref{fig:regions} has the highest
probability in the whole study region. This is exceptional as lower terrain
everywhere else has lower probabilities than higher terrain in its vicinity.
By mid-July lightning activity favors higher terrain. Lightning probability
again clearly differs between valleys and moutain ranges, however now reversed
from mid-May. The probability in valleys is considerably lower than over the
mountains. Overall, the strongest activity has shifted to the south of the
Alpine crest and probabilities even exceed the ones in the former hotspot in
the flat land in the vicinity of location F. One month later in mid-August the
overall pattern is the same but probabilities are lower---with the exception
of location I where the probabilities remain relatively high even into
mid-September, by which time lightning activity is low everywhere, most
pronouncedly so north of the Alps.

Contrary to the results from the GAM-method, results from the cell-count method
in Fig.~\ref{fig:maps}b are noisy and lack details  despite averaging the
results over an 11-day window centered on that particular day. Grid boxes with
probabilities of more than 10~\% occur next to boxes with 0~\%, which
necessitated a different color scale in the figure. Only the major features of
lower lightning probabilities/frequencies over the high mountains early in the
lightning season, the shift into the high mountains and south of the Alpine
crest, and the hotspot near location I (Fig.~\ref{fig:regions}) are visible,
admittedly more easily so when one has previously seen the results from the GAM-method.
Sample sizes in the grid boxes are simply far too small to obtain reliable
frequencies of occurrence of lightning, which is a rare event of only about
1~\% per day in a 1~square-kilometer grid box. The uncertainty of the
cell-count method can be determined using the modified Wilson method
\citep{brown2001} to estimate the width of the 95~\% confidence interval for a
binomial distribution (lightning yes/no) for a sample size of 11---the number
of years in our data set. The width is a staggering 27~\% for an event with a
probability of 1~\%. Even increasing the sample size by a factor of 11 to
$11 \times 11 = 121$ using a time-window of 11 days instead of 1 day as is done in
Fig.~\ref{fig:maps}b still yields a width of 4.9 percentage points. Therefore
finding a lot of noise in the results from cell-count method comes as no
surprise.

\begin{figure}[t!]
\centering
\includegraphics[width=10.7cm]{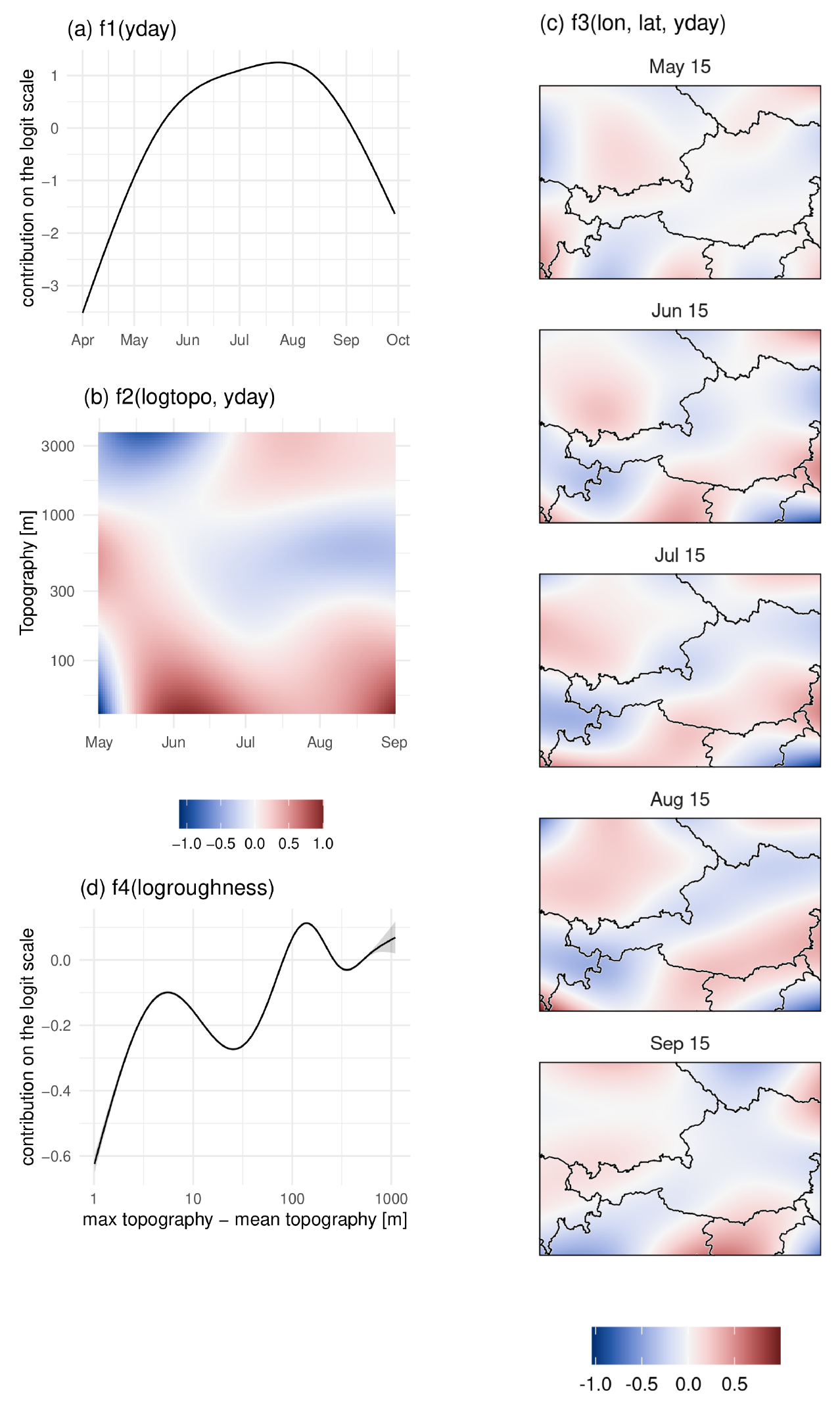}
\caption{Estimate effects of the GAM: a.\ the main seasonal cycle $f_1$,
         b.\ the correction of the seasonal cycle conditioned on the topography,
         c.\ the correction of the seasonal cycle given the geographic location,
         d.\ the effect of the roughness. All effects contribute to the additive
         predictor on the logit scale.}
\label{fig:effects}
\end{figure}

The power of the GAM-method and its ability to produce detailed and smooth
results even at such high spatial and temporal resolution lies in its
harnessing of information from grid boxes that share seasonal, topographical
and regional characteristics. Extracting the functional form of these
characteristics in turn leads to increased understanding of the contributors to
the overall spatio-temporal distribution of the rare event ``lightning
occurrence''. Fig.~\ref{fig:effects} shows the effects of each of the additive
terms in eq.~\ref{eq:gam} as they contribute to $\textrm{logit}(\pi)$.
Seasonally (Fig.~\ref{fig:effects}a), the
probability of lightning for the whole region increases rapidly in spring,
tapers off gently in autumn, and slowly increases during main lightning season.

The elevation effect (Fig.~\ref{fig:effects}b) varies seasonally. Note that the
effect is logarithmic (cf.\,eq.\,\ref{eq:gam}) but the ordinate axis has been
labeled in meters to ease interpretation. This effect shows a reduced
likelihood for lightning at the highest elevations early in the season, which
is reversed in the second half of June. The sign reverses last at the very
highest elevations, where snow melts last. Snow-covered ground greatly reduces
sensible heat flux and thus the chances for convection to occur. The seasonal
elevation effect also changes sign in elevations typical of intra-alpine
valleys. It is increased in spring, when the valleys are already free of snow
and reduced in summer and early autumn. The very lowest elevations, found
mostly at the southern edge of the study region, have an enhanced effect from
mid-May on.

The regional effect in Fig.~\ref{fig:effects}c accounts for adjustments needed
to add to the effects of season and elevation. At first glance, the patterns do
not seem to vary seasonally but a closer look reveals a shift of enhanced and
reduced regions from spring into summer and fall. In spring the lightning
probability in some parts of the lowlands and rims on either side of the Alps
and the high mountains in southeastern Switzerland is enhanced.
By mid-July lightning activity along the whole
southern side of the Alps is enhanced as well as in the westernmost and
easternmost parts of the lowlands north of the Alps. By mid-September, at the
end of convective season in most of the region, the northwestern rim and
lowlands and especially the southeastern rim and lowlands have enhanced
lightning probabilities.

Protruding topography such as peaks increases the likelihood of lightning.
Again, the effect is logarithmic but elevation difference in
Fig.~\ref{fig:effects}d is given in units of meters. Here the difference between
the maximum elevation at 90 m horizontal resolution to the average elevation in
the one-square-kilometer grid box serves as proxy for how jagged the terrain
is.  Overall, where that difference exceeds about 100~m, lightning likelihood
increases somewhat, which is in alignment with findings from lightning research
\citep[e.g.][]{rakov2003, kingfield2017, feudale2013}.

\begin{figure}[t!]
\centering
\includegraphics[width=8.3cm]{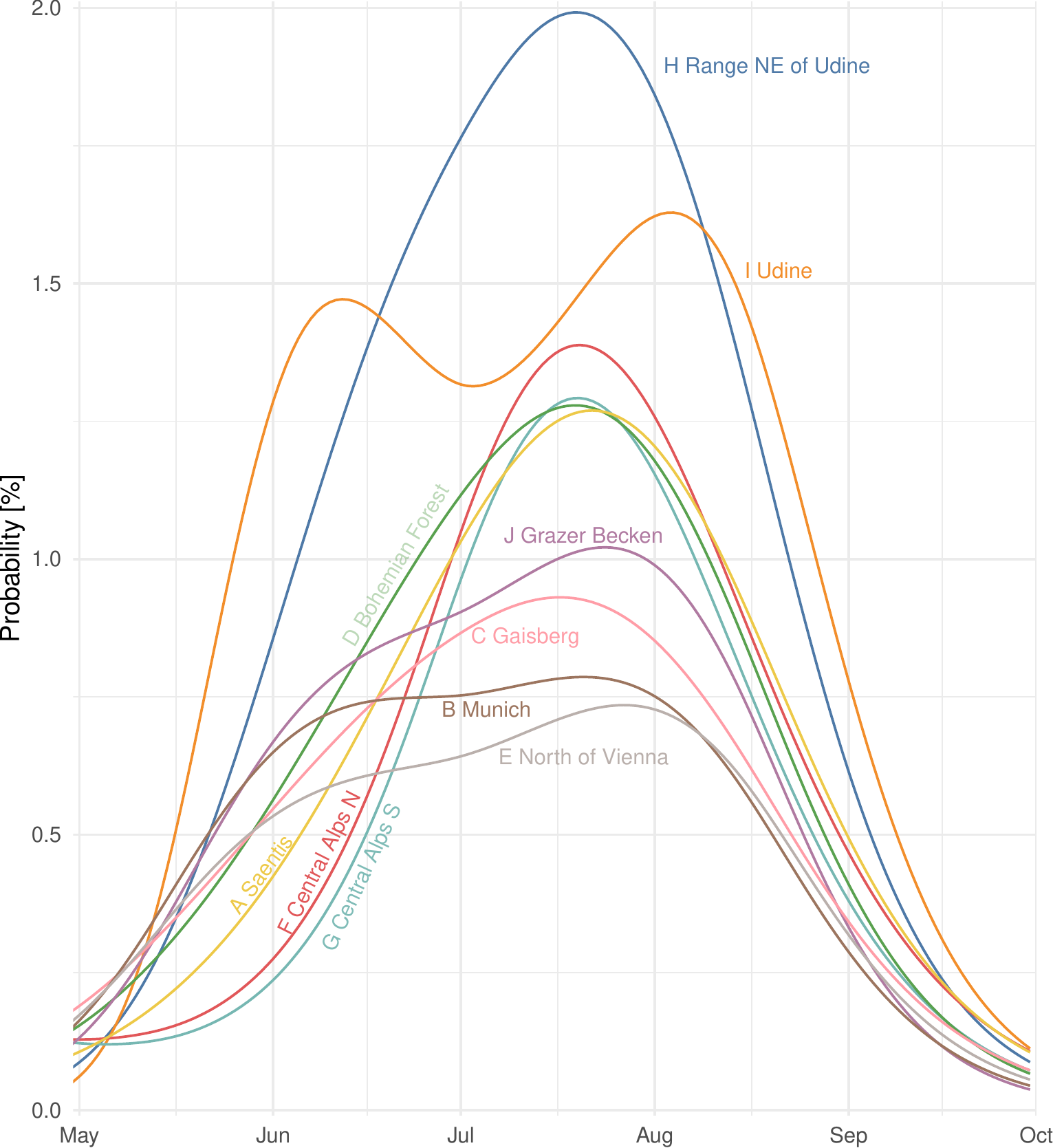}
\caption{Climatology cycles computed by the GAM for the locations from table~\ref{tab:locations}.}
\label{fig:cycles}
\end{figure}

Time series of lightning probability at specific locations in the eastern Alps
in Fig.~\ref{fig:cycles} exemplify the sum of the seasonally varying regional
and elevation effects and the roughness effect from Fig.~\ref{fig:effects} and
fill in the times between the regional maps of Fig.~\ref{fig:maps}. The lightning
season starts about three weeks later at the highest mountains, exemplified by
locations F and G, than at other locations (cf.\ Fig.~\ref{fig:regions} and
Table~\ref{tab:locations} for their coordinates). The difference in probability
between the two locations F and G stems from their difference in elevation.
Similar peak probabilities are reached by locations north of the Alps at A
(Saentis) and D in the Bohemian Forest, although they are considerably lower.
Their seasonal cycles are shifted by about one week, with D starting out one
week earlier and A lasting one week longer. Lightning season at location C
(Gaisberg), just south of D at the northern rim of the Alps starts about at the
same time but peak probabilities are about one third lower. Both A (Saentis)
and C (Gaisberg) are locations with a long history of direct lightning
measurements from instrumented towers. Although both are at the northern rim of
the Alps, seasonally varying regional differences (Fig.~\ref{fig:effects}c) and
an elevation difference of 1.1~km result in different shapes and peaks of their
seasonal lightning probability distributions. Flat terrain north of the Alps,
exemplified by locations B and E, experiences lightning less frequently than
the mountainous counterparts A, C and D. They also reach a first peak in
probability earlier and then increase slightly to a second peak at the end of
July. Locations in the flat lands south of the Alpine crest (E and I) show a
similar bimodal distribution. Location I, which is situated in a basin south of
the Alps has the most abrupt increase of lightning probability of all 10
locations shown. Despite its low elevation, its peak probabilities exceed the
ones of the highest mountains near the Alpine crest. It is located in the
hotspot so clearly visible in the regional probability maps in
Fig.~\ref{fig:maps}. However, probabilities at location H at the first mountain
range north of I are even higher.

\section{Discussion and Conclusions}\label{sec:discussion}

Lightning location systems (LLS) are unique among atmospheric measurement
systems in that they provide continuous measurements in both time and
space---two-dimensional or three-dimensional depending on LLS type. Since they measure
very rare events, translating these measurements into high-resolution
climatologies is difficult. Most climatologies so far have used the cell-count
method of counting how many flashes occurred within a particular time-space
cell, which necessitates making the spatial and/or temporal dimensions of such
a cell large in order to achieve a sufficient signal to noise ratio. Frequently
cell-sizes of 10--100 square kilometers by 1--12 months have been used
\citep[e.g.][]{schulz2005, feudale2013, poelman2016}. \citet{bourscheidt2014,
kingfield2017, simon2017spatio} demonstrated how the cell-count method is
unsuitable for smaller cell sizes corroborating the theoretical considerations
in \citet{diendorfer2008}, who used the Poisson distribution to derive the
minimum number of flashes in a cell required for reaching a specified accuracy.
The Poisson distribution is the limiting case of the binomial distribution,
which describes the binary event of the occurrence of lightning (yes/no). We
therefore use the binomial distribution to compute the 95~\% confidence
interval for a lightning probability of 1~\%. It has a width of staggering
27~\% for a sample size of 11 for a particular day in our 11-year data set.
Increasing the time dimension of a cell from 1 day to averaging over 11 days to
obtain a sample size of 121 still gives an uncertainty of 4.9~\%, five times
larger than what we need to detect. Fig.~\ref{fig:maps}b illustrates the high
noise level the cell-count method produces for high-resolution cells of
1~km$^2$ times 1 day. Fig.~\ref{fig:maps}a, on the other hand, produced with the
generalized additive model (GAM) method, is devoid of noise and yet provides
intricate spatial details for any given day in the lightning season (of which
the map shows five). Similarly spatially detailed climatological maps can be
achieved using a bivariate Gaussian error distribution for each lightning
discharge instead of a precise location \citep{bourscheidt2014, kingfield2017}.
The advantage of the GAM method is its ability to harness regional, elevational
and seasonal characteristics that cells share that need not be immediately
``next'' to another and thus make a high combined spatio-temporal resolution
possible.  To our knowledge, this paper presents the first daily-resolved
kilometer-scale lightning climatology over a large region. GAMs have another
advantage: the resulting functional forms of characteristics shared among cells.
Fig.~\ref{fig:effects} yield insights about the processes contributing to the
final climatological result.

The purely seasonal effect (Fig.~\ref{fig:effects}a) shows a rapid increase of ligthning probability in late spring but a more gradual tapering off in late summer and early fall, which was previously also found for a smaller region in the eastern Alps \citep{simon2017spatio}. The peak, however, is less sinusoidal but more plateau-like, probably due to the larger region where different subdomains peak at different times (cf. Fig.~\ref{fig:cycles}.

Lightning probability also varies with elevation (Fig.~\ref{fig:effects}b). \citet{smorgonskiy2013, simon2017spatio} found an overall increase with elevation. The GAM method allowed to implement a seasonally varying elevation effect and found an increased probability in spring at elevations in lowlands and intra-Alpine valleys compared to higher elevations (above approx. 1200~m~msl), which are still (partly) snow covered. This pattern reverses later and high elevations have a higher probability when the snow has melted. The most pronounced increase is at the very lowest elevations found on the southern side of the Alps (vicinity of location I) with maxima in June and late August/early September.

This regional hotspot was already known from earlier cell-count climatolgies
\citep[e.g.][]{schulz2005, feudale2013, taszarek2019}. With GAMs we could
additionally identify how the regional differences vary with season---in
extension to \citet{simon2017spatio}---when different elevations are already
accounted for.

The approach of using similar characteristics to compute high-resolution yet non-noisy lightning climatologies with GAMs pioneered by \citet{simon2017spatio} can be expanded by including further characteristics beyond region, elevation and time of year. Snow cover could be explicitly included since it suppresses convection. In our analysis it is implicitly contained in the seasonally varying elevation effect where lightning season at the highest elevations with the longest snow cover duration has a delayed start. Further characteristics of earth's surface such as slope angle and exposure, soil moisture, amount and type of vegetation cover, and land/water could be included as well as the presence of tall structures, which can trigger lightning \citep{rakov2003}. The GAM approach can also use further additive terms with output from numerical weather prediction models in order to provide \textit{forecasts} of the probability of thunderstorm occurrence \citep{simon2018}.

\section*{Computational Details}

The results in this study were achieved using \textsf{R}, a software
environment for statistical computing and graphics. The add-on packages
\textbf{mgcv} were used for building the statistical model.

\section*{Acknowledgements}

This work was funded by the Austrian Science Fund (FWF, grant no.~P\,31836) and
the Austrian Research Promotion Agency (FFG, grant no.~872656). Finally, we thank
Gerhard Diendorfer and Wolfgang Schulz from ALDIS for a fruitful and pleasant collaboration 
on lightning-related research and for providing the data for this study.

\section*{Statements and Declarations}

The authors declare that they have no conflicts of interest.

\bibliography{manuscript}

\end{document}